\documentclass[twocolumn,showpacs,amsmath,amssymb,amsfonts,aps,pre,superscriptaddress,longbibliography,10pt]{revtex4-1}

\usepackage{subfigure}
\usepackage{graphicx}
\usepackage{dcolumn}
\usepackage{bm}
\usepackage[english]{babel}
\usepackage{upgreek}
\usepackage{longtable}
\usepackage{epstopdf}

\usepackage{color}

\begin{document}

\title{Charging changes contact composition in binary sphere packings}

\author{Andr\'e Schella}
\email{andre.schella@ds.mpg.de}
\affiliation{Max Planck Institute for Dynamics and Self-Organization G\"ottingen, 37077 G\"ottingen, Germany}

\author{Simon Weis}
\affiliation{Institute of Theoretical Physics I, University Erlangen-N\"urnberg, Staudtstra{\ss}e 7, 91058 Erlangen, Germany}

\author{Matthias Schr\"oter}
\affiliation{Max Planck Institute for Dynamics and Self-Organization G\"ottingen, 37077 G\"ottingen, Germany}
\affiliation{Institute for Multiscale Simulation, Friedrich-Alexander University, 91052 Erlangen, Germany}

\date{\today}
\begin{abstract}
Equal volume mixtures of small and large polytetrafluorethylene (PTFE) spheres are shaken in an atmosphere of controlled humidity 
which allows to also control their tribo-charging. We find that the contact numbers are charge-dependent: as the charge density of 
the beads increases, the number of same-type contacts decreases and the number of opposite-type contacts increases.
This change is {\it not} caused by a global segregation of the sample. 
Hence, tribo-charging can be a way to tune the local composition of a granular material. 
\end{abstract}


\maketitle 

\section{Introduction}
\label{sec:introduction}
The term granular media comprises all ensembles of particles where the individual entities are large enough 
to be unaffected by Brownian motion. Besides gravity and contact forces, the dynamics of granular media
is also controlled by forces originating from the surface of the particles: electrostatic 
interactions \cite{Matsusaka2010}, capillary forces \cite{herminghaus_wet_2013}, 
and friction \cite{schroeter:17}. 
Understanding the role of these forces is not only an interesting scientific problem, but also important for technological 
applications because many raw materials in industry come in granular form~\cite{Duran2000}. 
Especially tribo-charging of granular particles proves to be challenging because it can lead to both
repulsive and attractive interactions between the 
beads~\cite{Lee2015,Qin2016,Chen2016a,Yoshimatsu2016,Kolehmainen2016b,Yoshimatsu2017}. 

The simplest model system to investigate the generally poly-disperse granular materials are binary sphere mixtures. 
They have been widely studied with respect to their jamming behavior~\cite{Hopkins2011,Hopkins2013,Chen2015,Koeze2016}, 
their structural features~\cite{Dijkstra1998,Kummerfeld2008} and their binary contact 
numbers~\cite{Epstein1962,Pinson1998,Biazzo2009,Meng2014,Kumar2016}. 
Binary sphere packings agitated vertically tend to 
segregate~\cite{Rosato1987,Knight1993,Kudrolli2004,Schroeter2006,Garzo2011,Brey2011}. 
Depending on the prevailing segregation mechanism, the larger spheres either rise to the top
(which is also called the Brazil nut effect), or they sink to the bottom.
Segregation is a common problem in the manufacturing industry where 
mixing of different types of granular materials is often a crucial process \cite{Ottino2000,Muzzio2002,Lu2005,Cheng2014}. 

Tribo-charging is pervasive in the handling of granular material because
every  time two materials get in contact some charge will be transferred~\cite{Harper1957,McCarty2008,Matsusaka2010}. 
Tribo-charging of granular samples can lead to the formation of clusters~\cite{Lee2015}, de-mixing \cite{Mehrota2007},  
or even prevent pore clogging~\cite{Chen2016b}. 
Recently we have shown that tribo-charging can also be used to counteract segregation \cite{Schella2017}. 

The amount of tribo-charging is known to depend on the humidity of the 
air~\cite{Nieh1988,Pence1994,Greason2000,Rhodes2003,Nemeth2003,Vandewalle2012,Xie2016,Schella2017}. 
Here, we use this dependence to control the amount of surface charges on the beads 
in a binary mixtures of Teflon spheres. At the same time we ensure that the charges are large
enough to avoid global segregation.
Using X-ray tomography we then investigate how the 
composition of small-small, small-large, and large-large contacts changes 
as a function of the surface charge.

\section{Experiment}\label{sec:experiment}
All experiments are performed with a mixtures of  approximately 10000 small 
and 1483 large  polytetrafluorethylene (PTFE) spheres, purchased from TIS. 
The radius of the small spheres $r_{s}$  is 0.795 \,mm  ($\pm$ 3.1\,\% according to the
manufacturer), the large spheres have a radius $r_{L}$ of 1.5\,mm ($\pm$ 0.8\,\%).

The binary mixtures are shaken sinusoidally in cylindrical 
containers (diameter 50\,mm, made of polyamide Nylon 6-6) 
using an electromagnetic shaker (LDS 406). In order to assure 
steady state conditions, all samples are shaken for one hour
at a frequency of 100\,Hz and an acceleration of 2\,g. 
To avoid the accumulation of dust, the beads and the container 
are cleaned with ethanol and pure water 
after each five measurements. 

The average charge of individual beads is measured after the shaking has stopped 
by extracting each ten large and small beads from the sample using an antistatic tweezer. 
The beads are then deposited into a Faraday Cup connected to a Keithley 6514 electrometer.
Because the magnitude of the charge on a dielectric particle will scale with the beads' surface area, we 
consider here the surface charge density $\sigma_{L,s} = Q_{L,s} / 4 \pi r_{L,s}^2$ of 
large resp.~small beads instead of the total magnitude of charge $ Q_{L,s}$~\cite{Chademartiri2012}. 
We note, that the sum of all charges on the beads is not necessarily zero, as the walls of the
shaking container will also charge electrostatically.

In order to modify the charge accumulation on the beads, the experiments are performed 
under controlled relative humidity (RH). A self-built climate chamber equipped with a cooling trap and 
an ultrasonic transducer allows to tune the ambient humidity in the range 
between 10\,\%RH and 100\,\%RH~\cite{Schella2017}. 
The humidity inside the chamber is logged constantly and changes on 
average about 2\,\%RH during the course of an experiment. 
Humidity control is started one hour prior to the the experiment in order
to equilibrate the water content on the surface of the beads and the container walls~\cite{Zitzler2002}. 

An advantage of using PTFE beads is their high contact angle with water 
(108$^{\circ}$ \cite{Kwok1999}) which prevents the formation of capillary bridges at high humidity levels.  
Consequentially, segregation due to capillary attraction~\cite{Samadani2000,Geromichalos2003}
will not affect our experiments. 

\begin{figure}[t]
\includegraphics[width=1\columnwidth]{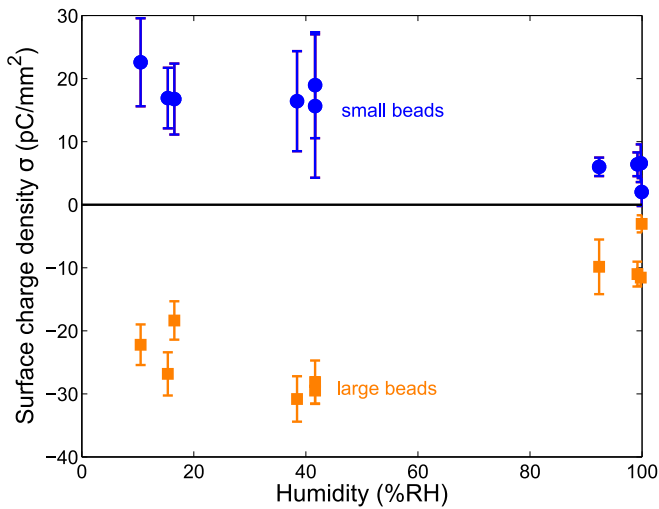}
	 \caption{The amount of charge accumulated by shaken PTFE spheres depends on the relative humidity,
           the sign depends on the size of the particles with small particles being positively and large particles being negatively charged. 
             Data are taken from samples of equal volumes of small and large PTFE spheres, shaken
             vertically in a polyamide container.
             }
	 \label{fig:charge_versus_RH}
\end{figure}

Figure \ref{fig:charge_versus_RH} demonstrates that under our shaking conditions large spheres charge negatively and small spheres charge positively. This observation is the opposite of what has been found in previous granular experiments~\cite{Hu2012,Xie2013,Waitukaitis2014} and predicted by some models of same-material tribo-charging~\cite{Kok2009,Lacks2011}. A result similar to our observation was found in experiments with spheres sliding along a plane made from the same material~\cite{Lowell1986a,Lowell1986b}.

To estimate the threshold for tribo-charging, we have de-ionized large PTFE beads on a grounded metal plate using 
an electrostatic ion gun prior to depositing them in the Faraday Cup. The residual charge density on these beads 
was found to be $\sigma_{th} = -1.8\,\mbox{pC}/\,\mbox{mm}^2$ ($Q_{th} = -52\,\mbox{pC}$), 
which is comparable to previous results~\cite{Kaufman2008}. 

\subsection{X-ray computed tomography}~\label{subsec:tomo}
The structure of the packings created by shaking
is analyzed using X-ray computed tomography. The tomography setup (Nanotom, General Electrics) is operated
 at 130\,kV and 90\,$\mu$A using a tungsten target. The side length of a voxel
(which is the 3D equivalent of a pixel)
 is 60$\mu$m$^3$ and data sets consist typically
 of $900 \times 900$ voxel in horizontal direction and, depending on the expansion of the bed,  800 to 900 voxel
 in vertical direction. 

Particle centers and radii are identified using the image processing procedure described in Ref.~\cite{Schella2017}. 
Since the structural features of the tribo-charged mixtures might be modified in the vicinity of the walls~\cite{Kumar2014},
we exclude all particles which are closer than three large particle diameter to the container walls from our further analysis.
At the top and bottom we discard two layers of large particles. The remaining core region consists of $3090\pm 430$ particles.

\begin{figure}[t]
\includegraphics[width=1\columnwidth]{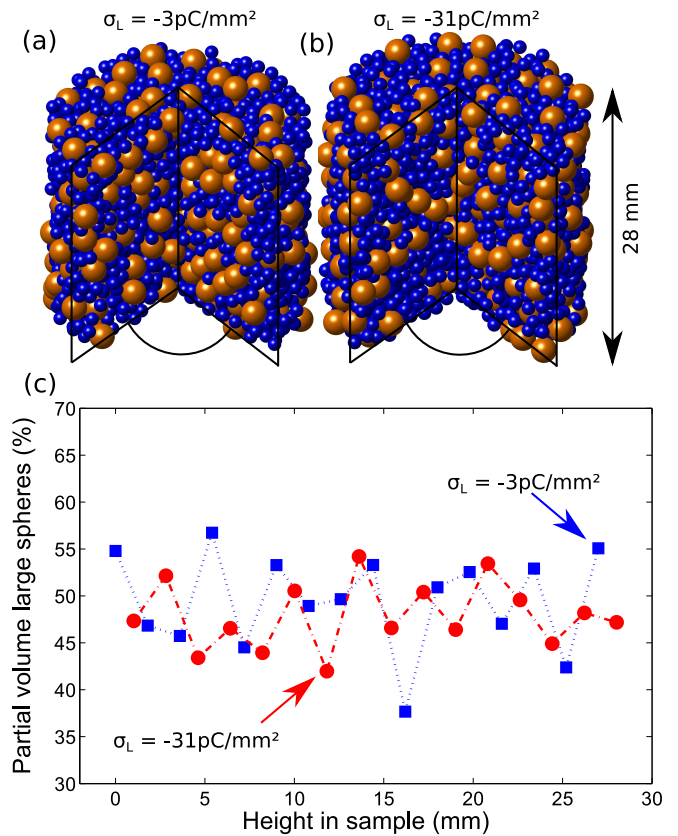}
	 \caption{Renderings of binary packings of small and large PTFE spheres.
          Particle positions were acquired using X-ray tomography. 
          A $90^{\circ}$ wedge has been removed to visualize the interior structure. Both samples have been 
          shaken vertically for one hour at 2$g$, but at different humidity
 levels. (a)  corresponds to the 
          sample with the least charged large spheres (b) to the one with strongest charge.
          Neither sample shows macroscopic segregation, i.e.~differences in the vertical  
          distribution of the large spheres with respect to the small ones. This is 
          also shown quantitatively in panel (c) which displays the height dependence of
         the volume contributed by the large spheres to the total particle volume for low (blue squares) and high (red circles) charge density.
         }
	 \label{fig:tomo_example}
\end{figure}

Figures~\ref{fig:tomo_example}~(a) and (b) show reconstructed sphere positions 
from the inner part of two samples, the two panels correspond 
to the samples with the smallest and largest surface charge density on the large spheres.
Neither packing shows signs of vertical segregation. This can also be seen in figure~\ref{fig:tomo_example}~(c): 
within fluctuations the contribution of the large spheres to the total volume is one half, independent of height.
This result holds also for all other experiments reported here.

The X-ray tomographies allow us to compute both the average number of contacts of the spheres and the
volume fraction $\phi$ of the packing.
A binary mixture has four different contact numbers: first the number of
contacts an average large spheres forms with other large spheres $Z_{LL}$, or with small spheres $Z_{Ls}$. 
Then, the number of contacts an average small sphere forms with large spheres $Z_{sL}$ (which is different from 
$Z_{Ls}$, cf. Sec.~\ref{sec:contacts}) and finally the number of contacts between small spheres $Z_{ss}$. 
We have measured those four numbers by adapting the contact number scaling function method described
in \cite{Aste2005,schaller:13,Weis2016}. Details can be found in the Appendix~\ref{app:contact_number_analysis}. 

In order to compute the volume fraction of the analyzed region, without the interference of any boundaries,
we first perform a set Voronoi tessellation of our sample which assigns each point of the interstitial space
between the particles to the sphere which surface is closest ~\cite{Schroeder-Turk2010,Weis2016}. The global packing fraction is then computed as
\begin{equation}
  \label{eq:voronoi}
  \phi = \frac{  \frac{4 \pi}{3}  (N_L r_{L}^3 + N_s r_{s}^3)} 
              {  \displaystyle\sum_i^{N_L}{\nu_{L}^i} + \sum_j^{N_s}{\nu_{s}^j }}
\end{equation}
The enumerator contains the total volume of all the $N_L$ large and  $N_s$ small spheres
in the analyzed region and the denominator the sum of all the individual 
Voronoi volumes $\nu_{L}$ and $\nu_{s}$  of the large respectively small spheres.

\section{Charge controls the contact numbers}
\label{sec:contacts}
\begin{figure}[t]
	\centering
\includegraphics[width=0.98\columnwidth]{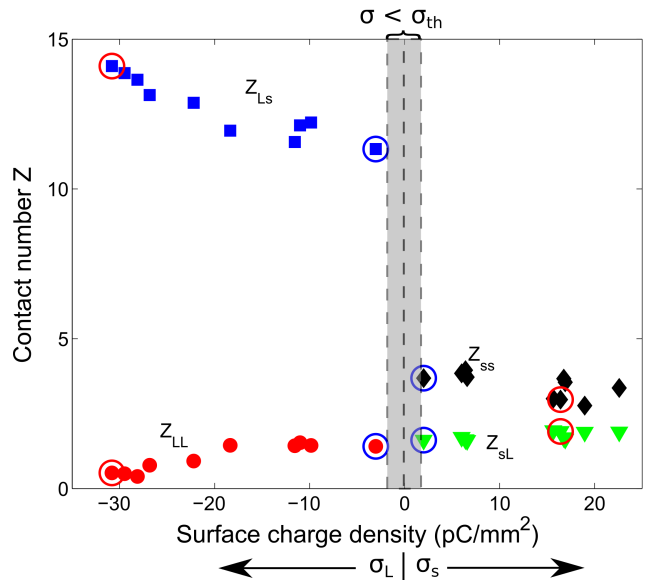}
	 \caption{Number of large-large $Z_{LL}$, large-small $Z_{Ls}$, small-large $Z_{sL}$ and small-small $Z_{ss}$ 
contacts in a binary mixture as a function of the average surface charge density of the large resp.~small beads. 
Circles identify the two packings depicted in Figs.~\ref{fig:tomo_example} a (blue) and b (red).
The shaded region corresponds to the residual charge regime where $|\sigma| < 1.8\,\mbox{pC}/\mbox{mm}^2$. }%
	 \label{fig:mPTFE_contacts}%
\end{figure}

Figure \ref{fig:mPTFE_contacts} shows the main result of our study:
the binary contact numbers exhibit a clear dependence on $\sigma_{L}$ and $\sigma_{s}$.
The numbers of large-small and small-large contacts, $Z_{Ls}$ and $Z_{sL}$,
 increases linearly with increasing  electrostatic charge density. 
At the same time the number of same type contacts, $Z_{LL}$ and  (less obvious) $Z_{ss}$,
decreases with increasing surface charge density.
This change in contact numbers is in good agreement with a simple model assuming that
like-charged large beads repel each other whereas oppositely charged particles attract each other. 

The increase of opposite type contacts in charged samples 
is also compatible with the visual impression gathered from 
figures \ref{fig:tomo_example} (a) and (b). While neither of the two packings
shows macroscopic segregation, the local structure differs in that the large particles
form more string-like structures in the highly charges sample. Similar structures have been 
identified in simulations of charged binary colloidal aggregates \cite{Barros2014} 
and mono-disperse charged grains~\cite{Chen2016a}. An interesting follow-up 
question will be if these changes in microstructure do also alter the macroscopic 
mechanical behavior of the material. This would open an avenue for granular packings with 
tunable properties.

There is a noticeable difference in how strong $Z_{Ls}$ and $Z_{sL}$  dependent on 
their respective $\sigma$, i. e.~$\partial Z_{Ls} / \partial \sigma_L > \partial Z_{sL} / \partial \sigma_s $. 
This difference can be explained using the fact that the total number of large-small
 contacts in a given volume is the same as the number of all small-large contacts:
$N_L Z_{Ls} = N_s Z_{sL}$.  Taking the derivative with respect to an average $\sigma$  we obtain
\begin{equation}
\label{eq:num_contacts}
\frac{\partial Z_{Ls}}{\partial \sigma} = \frac{N_s}{N_L}\frac{\partial Z_{sL}}{\partial \sigma}\qquad\mbox{.}
\end{equation}
where we have used the additional condition that the total number of particles in the observation volume 
is independent of the charge, which is indeed justified in our experiments.  
As we have studied equal volume mixtures, equation \ref{eq:num_contacts} 
predicts $N_s/N_L = (r_L /r_s)^3 \approx 6.7$.
A linear fit to the data of Fig.~\ref{fig:mPTFE_contacts} yields
 $\partial Z_{Ls}/\partial \sigma \approx 5.5 \; \partial Z_{sL}/\partial \sigma$ which is in reasonable 
agreement with the predicted slope ratio. 

We can also compare our contact numbers results with previous experimental ~\cite{Pinson1998} and theoretical
\cite{Biazzo2009,Meng2014} work on the contact numbers of uncharged binary mixtures.
A linear regression of our data and an extrapolation to the value $\sigma \rightarrow 0~\mbox{pC}/\mbox{mm}^2$ 
yields $Z_{LL}$ = 1.9, $Z_{Ls}$ = 10.9, $Z_{sL}$ = 1.6
and $Z_{ss} = 4$, which agrees well with the previously published results
for packings of comparable size ratio.

\section{Average contact number and global packing fraction}
\label{sec:AveZ_phiGl}
\begin{figure}[t]
	\centering
\includegraphics[width=1\columnwidth]{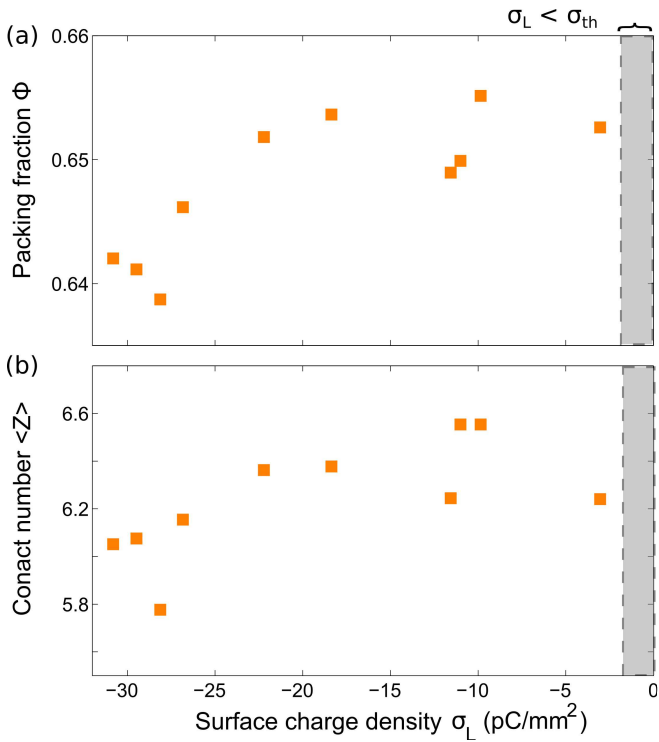}
	 \caption{The global packing fraction $\phi$ and the average contact number $\left\langle Z\right\rangle$ 
           depend weakly on the surface charge density $\sigma_L$.
           The shaded region marks the residual charge
           regime where $\sigma_L < - \sigma_{th}$.}
	 \label{fig:Zave_phiGl}
\end{figure}

In the previous section we have shown that tribo-charging leads to a local rearrangement and 
hence changes the binary contact composition. However, tribo-charging does also affect 
global quantities of the binary sphere packings, as shown in figure \ref{fig:Zave_phiGl}.
The packing fraction $\phi$ decrease approximately 1\% with increasing surface charge density. 
This trend is in agreement with simulations of mono-disperse particles \cite{Chen2016a}. 

The average contact number $\left\langle Z \right\rangle$ does also decrease with increasing 
surface charge density. Hence, the bed expands and gets looser. Such a correlation of $\phi$ and $\left\langle Z \right\rangle$ is to be expected based on previous studies of mono-disperse sphere packings \cite{Aste2005,schaller:15}.

Qualitatively, increasing the charge density on the beads will also increase attractive interactions between large and small particles. Thus, a decreasing packing fraction with increasing charge density seems counter-intuitive at first glance. However, attractive interactions also alter the mechanical stability of granular packings since these have a stabilizing effect, causing the formation of chain-like, porous structures~\cite{Barros2014,Chen2016a}. To what extend additional many-body~\cite{Cheng2014} or polarization effects of the dielectric beads~\cite{Barros2014,Lee2015,Qin2016} contribute to our findings has to be clarified in future studies.

\section{Summary}\label{sec:summary}
Binary systems of dielectric particles have been shaken vertically at different humidity levels which allows to 
control the tribo-charging of the beads. Because small and large beads differ in the sign of their charge, 
the resulting attractive interactions inhibit macroscopic segregation of the sample. At the same time the 
electrostatic interactions change the local structure of the packing: the stronger the charge carried by the individual 
particles is, the more likely becomes the formation of contacts between small and large beads at the expense of
same bead type contacts. Previous studies of binary packings stated that the composition of 
contacts can be changed by changing the number ratio of small to large particles. 
Here, we suggest an alternative route: The composition of contacts can also be altered  
by tribo-charging the particles.

\begin{acknowledgments}
We thank Wolf Keiderling for repeated mechanical support. 
\end{acknowledgments}

\appendix*\label{app:contact_number_analysis}
\section{Contact number analysis}
A contact between bead $A$ and $B$ is defined as touching beads,
i.e.~when the distance $d$ between the
beads is equal to the particle radii $d = r_A +r_B$.
Applying this definition to experimental data,
as e.g.~gathered by X-ray tomography, is a challenging task due to two
reasons. First,
errors in the image acquisition and processing add random noise to the
particle coordinates and therefore
distances between pairs of particles. And secondly, all granular particles
are to some degree poly-disperse, hence $r_A +r_B$ is not a constant but
depends on the individual particles
under consideration.
To mitigate these two problems we use an ensemble based fitting method
which determines
$Z_{AB}$ by modeling the effect of inaccuracies in the particle
coordinates using
the best average representations of $r_A$ and  $r_B$
\cite{Aste2005,schaller:13,Weis2016}.

The method works in two steps. First, the average
interparticles distance  $\langle r_A +r_B \rangle$ is determined from
the first
peak of the binary radial distribution function	$g_{AB}(d)$ which measures
the probability to find a particle of type $B$ in a distance $d$ from a
given particle of type $A$.
$g_{AB}(d)$ can be computed by counting the number of particles in spherical
shells around a reference particle:

\begin{equation}\label{eq:radial_distr_func}
	g_{AB}(d) = \left\langle \frac{1}{4\pi d^2 \rho} \sum_{B,j}{\delta (d -
| \vec{x}_{A} - \vec{x}_{B,j}| )} \right \rangle_{A}
\end{equation}
Here the sum over $j$ runs over all particles of type $B$ and the delta
function gives only a contribution if the
distance between the two particle centers  $| \vec{x}_{A} -
\vec{x}_{B,j}|$ is equal to $d$.
The triangular brackets denote the average over all particles of type $A$.
The  normalisation consists of two parts:  the  volume of the spherical
shell analyzed grows
with $4 \pi d^2 $ and by dividing with
the number density $\rho$ we assure that an uncorrelated system will
have $g_{AB}(d) = 1$.

\begin{figure}[t]
	\centering
\includegraphics[width=1\columnwidth]{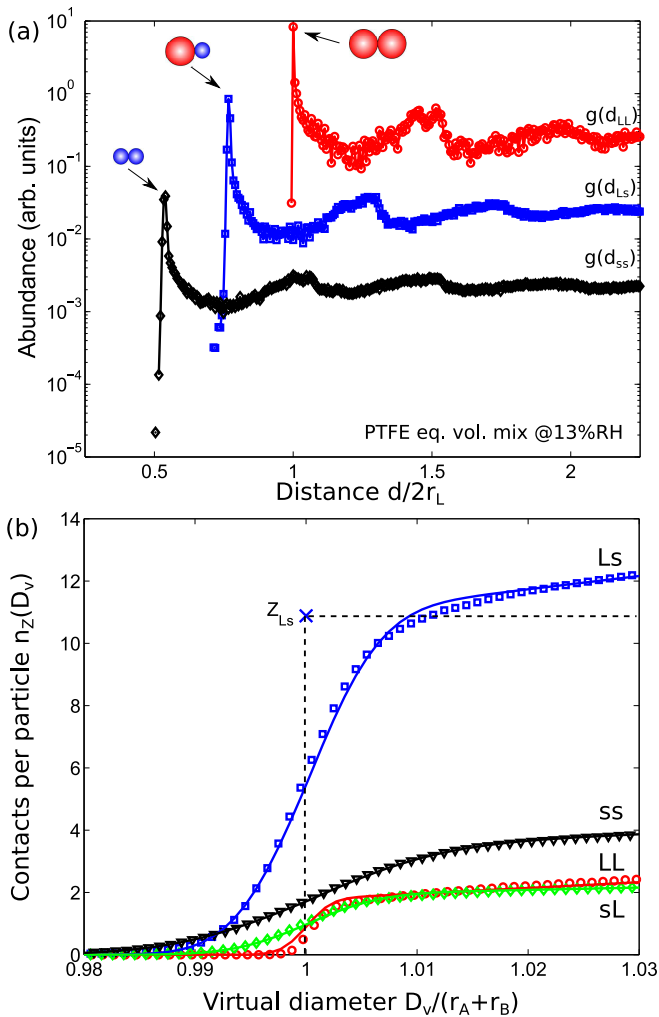}
	 \caption{(a) Binary radial distribution functions $g_{AB}(d)$ of an
equal volume mixture of tribo-charged
 PTFE spheres. The position of the first peak provides the best
estimates for the three different sums of radii.
Functions are shifted vertically for better visibility.
(b) In order to account for experimental uncertainties of the detected
particle positions, particle radii are scaled up
and down and the number of contacts per particle $n_Z$ is counted for
the different virtual diameters $V_v$.
Fitting this data with the contact number
scaling function Eq.~\eqref{eq:Aste_contact_model_full} allows us to
measure the three different contact numbers $Z_{AB}$.
}
	 \label{fig:gr_and_contacts_example}
\end{figure}

Figure~\ref{fig:gr_and_contacts_example} (a) shows the large-large
$g_{LL}(d)$, large-small $g_{Ls}(d)$
and small-small $g_{ss}(d)$  pair distributions for a mixture shaken at
approximately 13 \% RH.
The first peak in these distributions originates from particle pairs in
contact, therefore an extrapolation
of the peak positions provides the best possible estimate for the three
different combinations of
$\langle r_A +r_B \rangle$.

To determine the four different contact numbers $Z_{AB}$ we follow
an adapted version of the procedure described in Refs.~\cite{schaller:13,Weis2016}.
First we determine how the number of contacts $n_{Z_{AB}}$ (defined as
touching or overlapping particles)
changes if we multiply the particle radii with a scaling factor, thereby
creating particles with
virtual diameters $D_v$ ranging from 0.98 to 1.03 times $2 r_A$
respectively $2r_B$.
The resulting $n_{Z_{AB}}(D_v)$ can be seen in figure
\ref{fig:gr_and_contacts_example} (b).

The idea is that errors in the particle positions due to image
processing and polydispersity
should be Gaussian distributed. We expect therefore that for virtual
diameters smaller than
$D_{avg} = \langle r_A +r_B \rangle$ the
binary contact numbers $n_{Z}$ will follow a cumulative normal distribution
\begin{equation}\label{eq:Aste_contact_model}
	n_{Z}(D_v) = \frac{Z_{AB}}{\sqrt{2 \pi}\sigma}
        \int_{0}^{D_v}{\exp \left( - \frac{\left(D_v' - D_{avg}
\right)^2}{2\sigma^2}\right) d D_v'}
\end{equation}
where the experimental uncertainties are captured by the variance
$\sigma$ and $Z_{AB}$ is the average contact number
we try to determine.

For $D_v > D_{avg}$, a linear term has to be added to $n_{Z}(D_v)$ to
account
for close, but non-contacting particles, i. e.~particles from the right
shoulder of the first peak of
$g_{AB}(d)$.  The full contact number scaling function is thus given by
\begin{equation}\label{eq:Aste_contact_model_full}
	n_{CNS}(D_v) = n_{Z}(D_v)  +  \Theta (D_v- D_{avg})  \,m (D_v- D_{avg})
\end{equation}
with $m$ being an unknown slope and $\Theta$ the Heaviside function.

Figure \ref{fig:gr_and_contacts_example} (b) shows that equation
\ref{eq:Aste_contact_model_full}
provides reasonable fits for all four possible combinations of binary
contacts.
\bibliography{mybib_3Dbinary_clean}

\end{document}